# High-speed RF Switch Electronics for picking up of Electron–Positron Beam Bunches

Liujiang Yan, Lei Zhao, *Member, IEEE*, Jinxin Liu, Ruoshi Dong, Zouyi Jiang, Shubin Liu, *Member, IEEE*, Qi An, *Member, IEEE*

***Abstract*—Beam diagnostics is important to guarantee good quality of beam in particle accelerator. Both the electron and positron run in the tunnel in some modern electron positron colliders such as Circular Electron Positron Collider (CEPC) to be built and Beijing Electron Positron Collider II (BEPC II). To measure the electron and positron beams, picking up of these two different bunches in real time is of notable concern. Because the time interval between adjacent electron and positron bunches is quite small, for example, 6 ns in CEPC, high-speed switch electronics is required. This paper presents the prototype design of a high-speed radio frequency (RF) electronics that can pick up nanosecond positron–electron beam bunches with a switching time of less than 6 ns. Fast separation of electron and positron is achieved based on RF switches and precise delay adjustment of the controlling signals (~10 ps). Initial tests have been conducted in the laboratory to evaluate the performance of electronics; the results indicate that this circuit can successfully pick up the bunch signal within a time interval of 6 ns, which makes it possible to further measure the electron and position beams simultaneously.**

## I. Introduction

BEAM diagnostics is important to guarantee good beam quality, for instance, beam phase, position measurement [1–5], or bunch-by-bunch feedback system [7–9]. In some modern colliders, electron–positron beams follow an orbit in the ring and are detected by the detector [10–14]; then, output signals from the detector are recorded as electron–positron bunches. Therefore, a picking-up electronics is needed for the data acquisition system, considering the electron and positron signals are separately processed and analyzed [17].

A high-speed radio frequency (RF) switch electronics was designed to receive four channels of signal from the Beam Position Monitor (BPM) electrodes and one channel of LVPECL trigger pulse signal from the clock. Each BPM signal is detected as a nanosecond bipolar pulse. Then, each bipolar BPM signal is divided into two unipolar output signals for a positron beam bunch and an electron beam bunch, as shown in Fig.1. The best moment of segmentation is marked by the rising edge of trigger pulse signal. Finally, the output should be four channels of positron sensing signals, four channels of electron sensing signals, and a channel of sensing strobe signal.

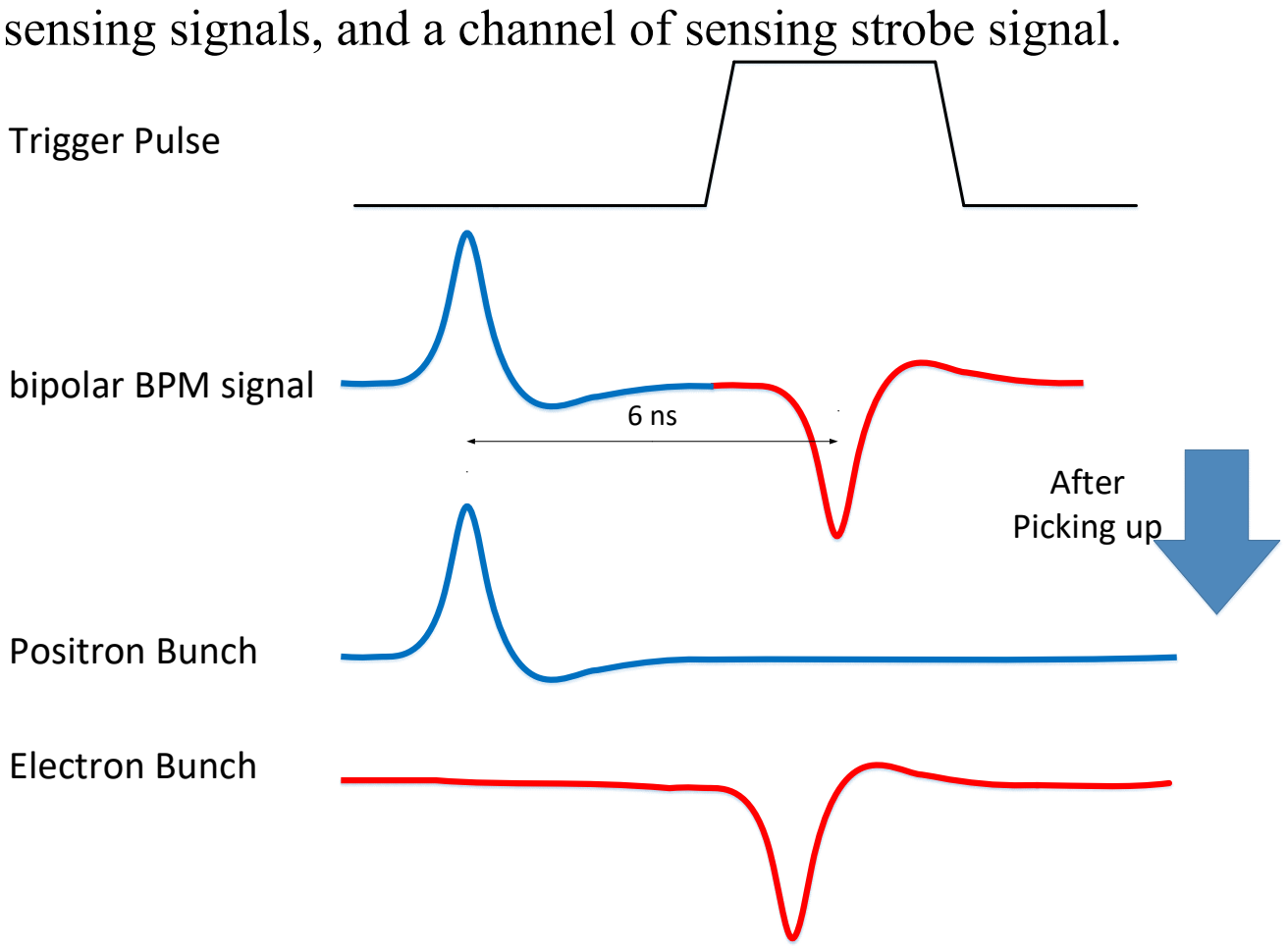

Fig.1 Principle of positron–electron beam bunches picking-up (1 channel)

The pulse width is 1 ns, and the interval time between the electron–positron bunches is also in the nanosecond scale, for example, 6 ns in the Circular Electron Positron Collider (CEPC) [18] or even less for some colliders. This is a challenge for our design.

## II. Hardware Design

### A. RF Switch Chips

To divide the bipolar BPM signal into two unipolar signals, the basic idea is to apply high-speed RF switch chips. Three versions of RF switch chip were used in this study: HMC427LP3, HMC347C8, and HMC347ALP3 [19–21]. The parameters are shown in Table I.

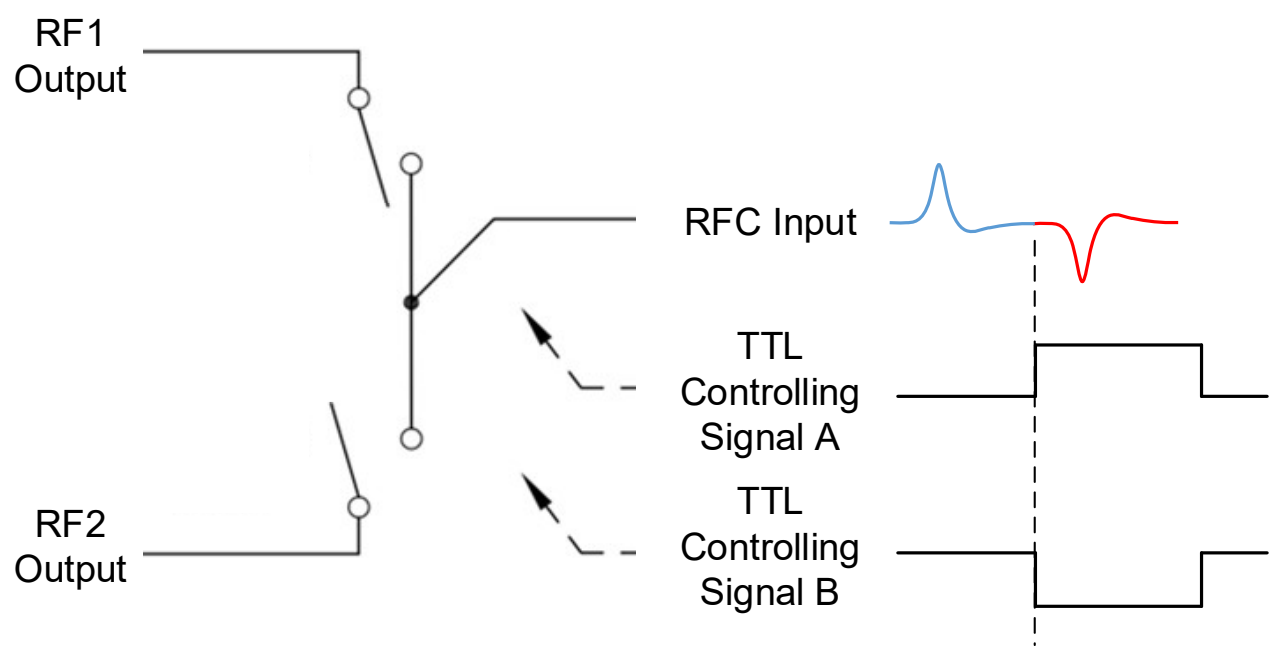

Fig. 2 Functional diagram of RF switch chips.

Manuscript received Jun. 24, 2018. This work was supported in part by the National Natural Science Foundation of China (11205153), in part by the Knowledge Innovation Program of the Chinese Academy of Sciences under Grant KJCX2-YW-N27, and in part by the CAS Center for Excellence in Particle Physics (CCEPP).

The authors are with the State Key Laboratory of Particle Detection and Electronics, University of Science and Technology of China, Hefei 230026, China, and Department of Modern Physics, University of Science and Technology of China, Hefei 230026, China (telephone: 086-0551-63607746, corresponding author: Lei Zhao, e-mail: zlei@ustc.edu.cn).

TABLE I
PARAMETERS OF THE RF SWITCHES

| Model Name | Insertion Loss | Isolation | tRISE, tFALL (10%/90% RF) |
|---|---|---|---|
| HMC347ALP3 | 1.6 dB @ 6 GHz | 46 dB @ 6 GHz | 3 ns @ 14 GHz |
| HMC427LP3 | 1.2 dB @ 6 GHz | 38 dB @ 6 GHz | 2 ns @ 8 GHz |
| HMC347C8 | 2.0 dB @ 6 GHz | 40 dB @ 6 GHz | 3 ns @ 8 GHz |

TABLE II
TRUTH TABLE OF THE RF SWITCHES

| Control Input | | Signal Path State | |
|---|---|---|---|
| A | B | RFC to RF1 | RFC to RF2 |
| High | Low | On | Off |
| Low | High | Off | On |

The simulation of three RF switch chips was conducted. The simulation results are shown in Fig. 3 and Fig. 4.

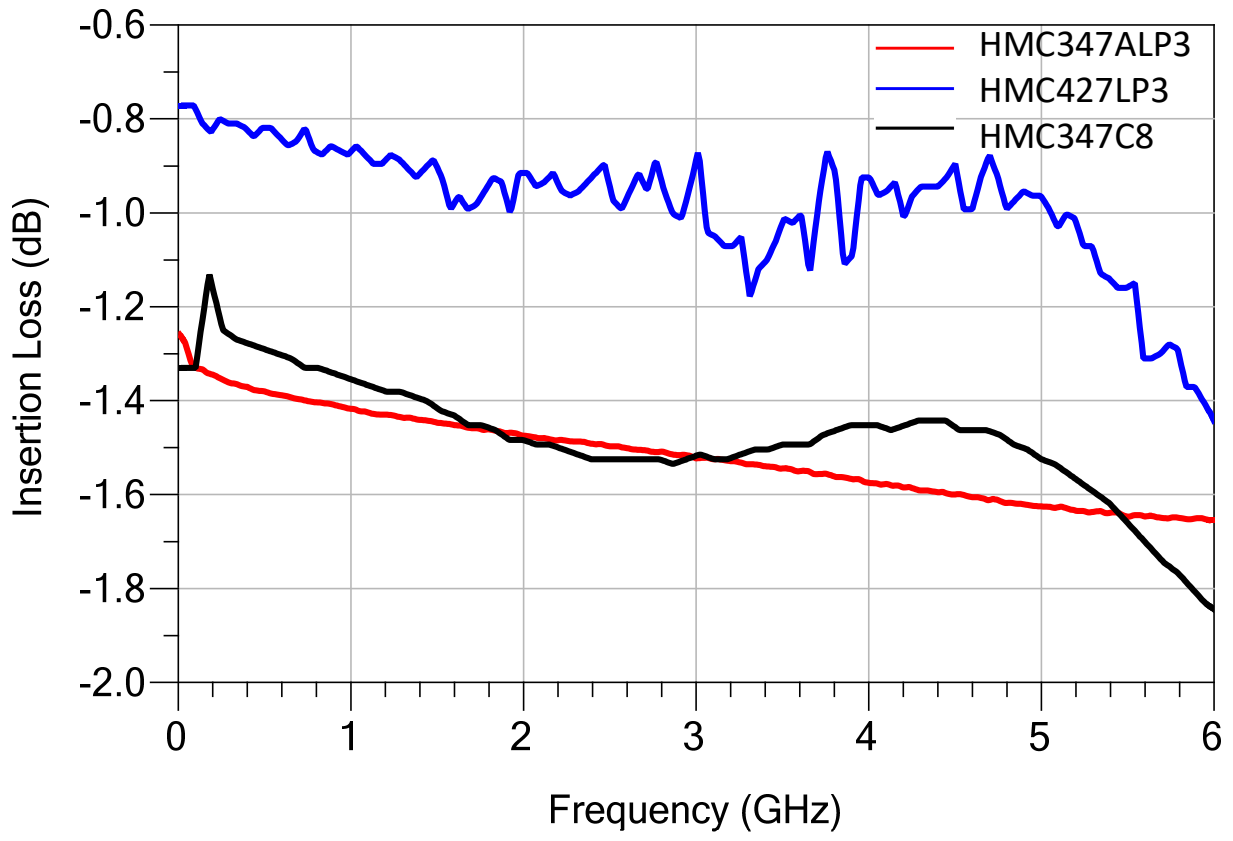

Fig. 3 Insertion loss simulation result.

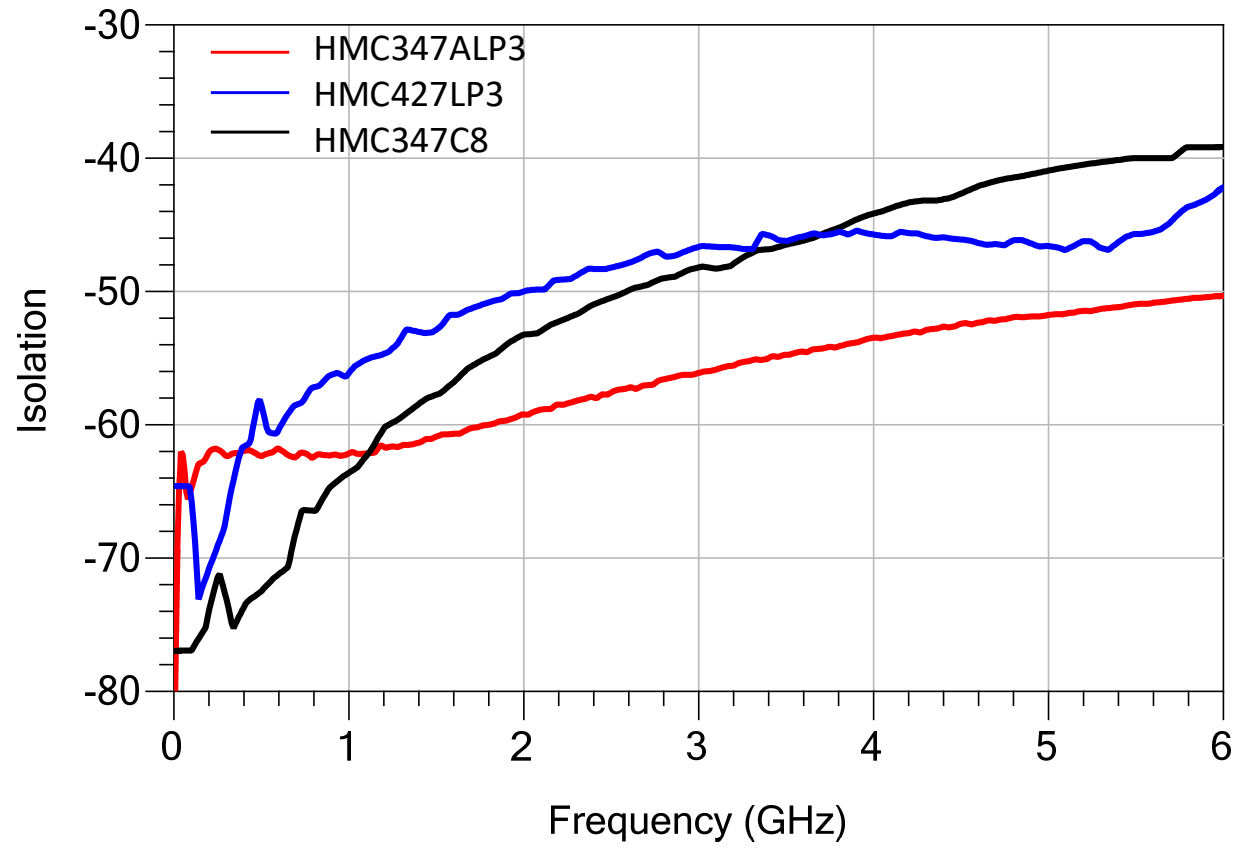

Fig. 4 Isolation simulation result.

According to the simulation results, three RF switch chips have an insertion loss better than −2 dB and an isolation better than −35 dB, whereas the signal frequency is below 6 GHz. The simulation results indicate that all the RF switch chips satisfy the requirements; HMC347ALP3 chip has a better and more reliable performance. Then, three verification electronics with a similar structure but different chips were designed and tested to estimate the function and performance of RF switch chips. Based on the abovementioned comparative analysis, it is possible to achieve higher performances. The hardware design is shown is Fig. 5.

## *B. Hardware Structure*

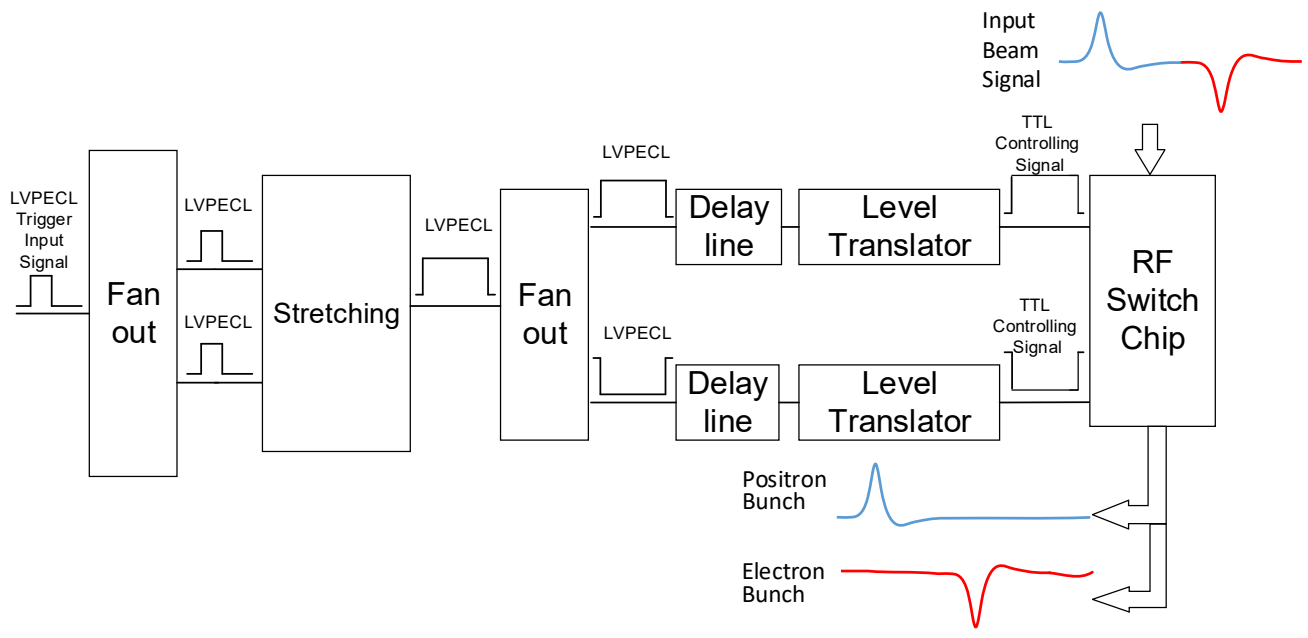

Fig. 5 Design of RF switch circuit.

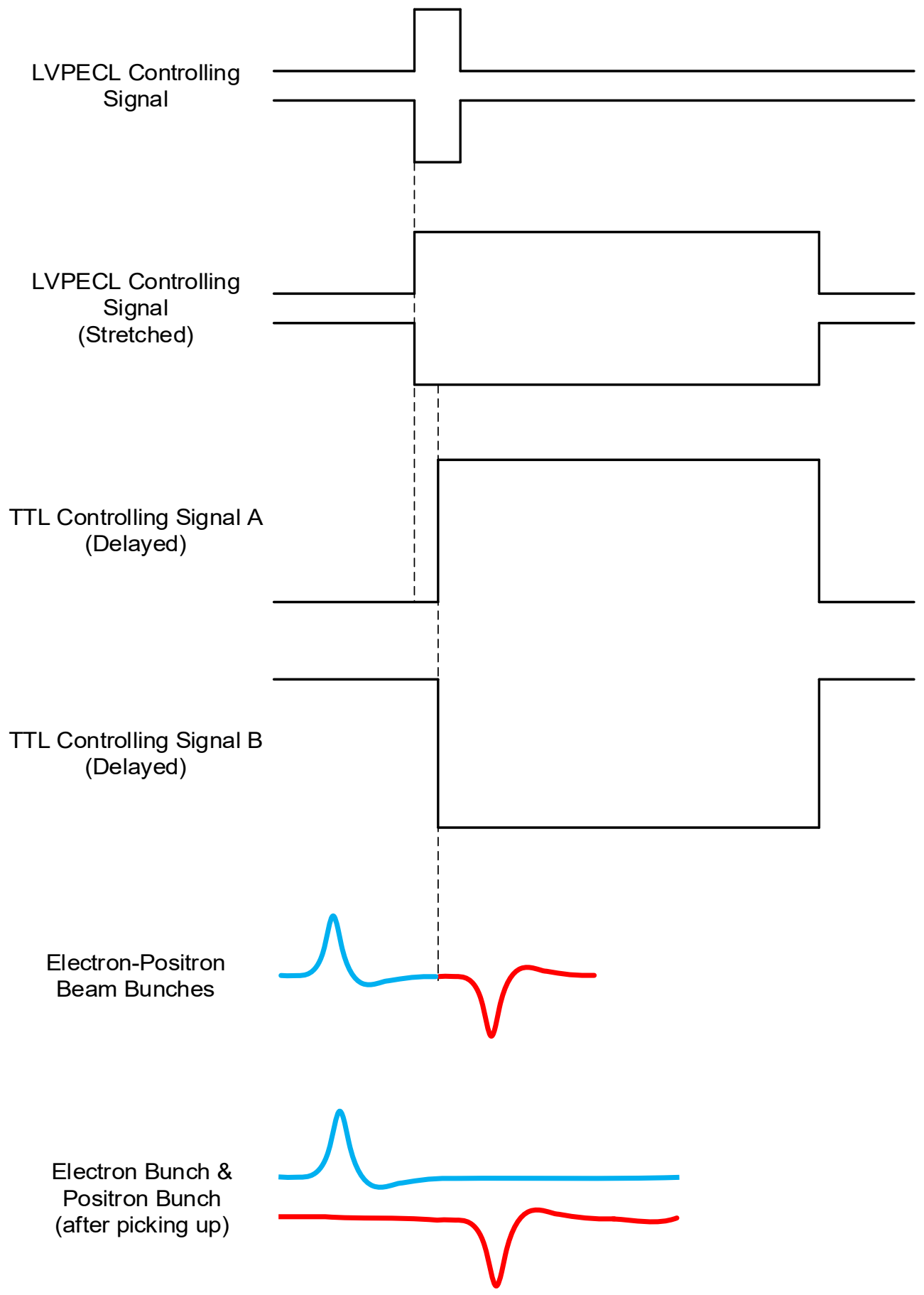

Fig. 6 Waveforms of signals in RF switch circuit.

In the design of RF switch electronics, several difficulties were considered. First, both the trigger pulse and bipolar BPM signal are narrow. To reach the proper switching moment, it is necessary to apply a finely adjustable delay line part. Second, the switch circuit requires high speed and low jitter; therefore, the input trigger pulse is differential LVPECL signal. Then, the RF switch chips are controlled by two single-ended TTL signals unlike the differential LVPECL input. Therefore, not only is a level translator part is applied, but also the LVPECL trigger pulse is divided into two channels through a fan-out circuit as well. The waveforms of signals in RF switch circuit have been shown in Fig. 6.

### C. *Fanning-out and Stretching Part*

The fan-out chip is MC10EP11, a differential 1:2 fan-out buffer with a typical frequency of more than 3 GHz and a typical RMS clock jitter of 0.2 ps, satisfying the design requirement.

According to the truth table of RF switch chip (shown in Table II), the best switching time is marked by the rising edge of trigger pulse. To be more specific, the switch circuit works normally when the trigger pulse is at a high level. Therefore, the controlling signals should be wide enough to cover the second half of the bunch. Because the trigger pulse is so narrow compared with the switching time, a stretching part was designed to guide the segmentation of BPM bipolar signal.

The stretching circuit is designed based on a D flip-flop and a counter, as shown in Fig. 7. MC10/100EP51 is applied as a differential clock D flip‒flop with a typical RMS cycle-to-cycle jitter of 0.2 ps and asynchronous reset.

The counter used is MC100EP016A, which is a high-speed pre-settable, cascade-able, 8-bit binary counter having an asynchronous reset. It has the maximum counting frequency of 1.5 GHz and the typical RMS clock random jitter of 2.6 ps.

A 100 MHz crystal oscillator is connected to the counter to obtain clock signals. In order to output 100MHz LVPECL signals, SiT9121 is chosen, which has a frequency stability of ±10 ppm.

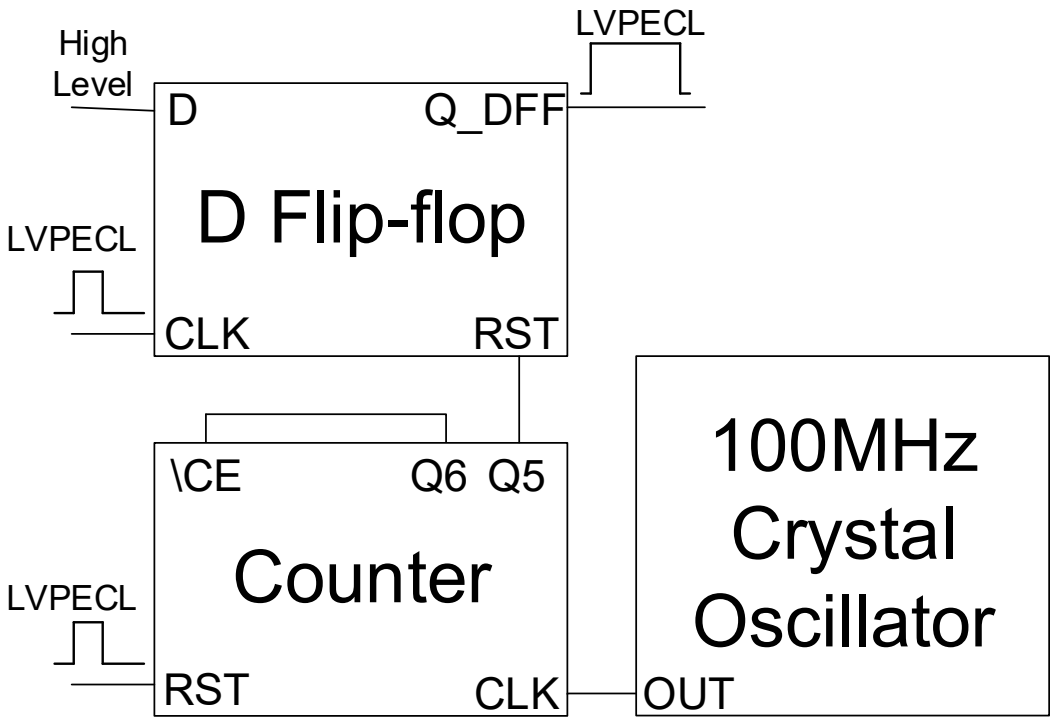

Fig. 7 Schematic of the stretching circuit.

The signal waveforms of the stretching part in time domain are shown in Fig. 8. After passing a fan out chip, the LVPECL trigger pulse is divided into two channels, and enters the CLK pin of the D flip-flop and the RST pin of the counter. The D-input of the D flip-flop is set to a fixed level of 2.3V. When the rising edge of the LVPECL trigger pulse enters the clock pin of the D flip-flop, the Q_DFF output becomes high. The output of Q5 pin is a square wave signal having a period of 640 ns and is connected to the RST pin of the D flip-flop. Due to this reason, the Q_DFF output turns out as low as 320 ns. As a result, the trigger pulse is stretched into a square wave having a width of 320 ns. In order to keep the circuit responding normally to input, the Q6 pin is connected to the Count Enable Control Input pin \CE.

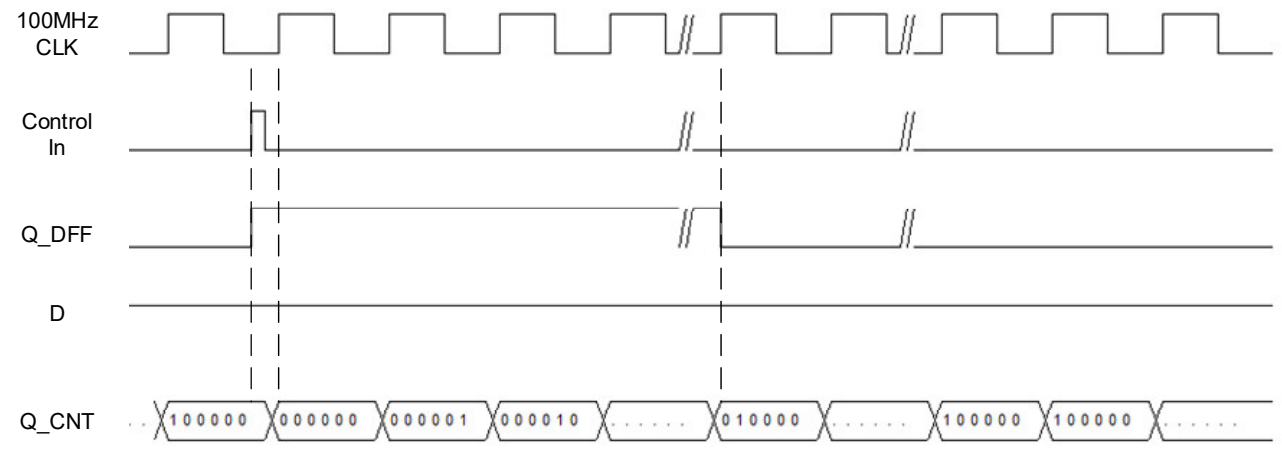

Fig. 8 Signal waveforms in time domain of the stretching part.

### D. *Delay Line and Level Translator Part*

When the electron-positron beam bunches are being picked up, it should be guaranteed that the controlling signals and the center of the bipolar input beam signal reach the RF switch chip at the same time. Therefore, a delay line circuit is employed to adjust the optimal switching moment of the RF switch chip. Furthermore, considering the truth table of the RF switch chip, which requires differential TTL input to suppress the electromagnetic interference, the delay time of controlling signals must be adjustable to guarantee that the controlling signals are synchronized precisely. It should be noticed that even a minor misalignment will result in a reduced performance of the switch electronics.

In this paper, the delay line is chosen to be SY89295U, which is a programmable delay line through a digital control signal. The delay line has the maximum operating frequency of 1.5 GHz and the delay time Integral Non-Linearity (INL) of ± 10 ps.

All RF switch chips work at TTL level. However, the trigger pulse is at the level of LVPECL for high speed and low jitter. In order to solve this discrepancy, a level conversion part is added to guarantee that it works precisely.

## III. Test Results

### A. *Testing System Setup*

Initial testing has been conducted to evaluate the working condition and performance of the switch electronics.

The testing system consists of an oscilloscope, a signal source, a DC power supply and an RF step attenuator. Fig. 9 shows the experimental setup during testing in the laboratory environment. Both the bipolar signal and the trigger pulse are generated by programming the 81180A signal generator.

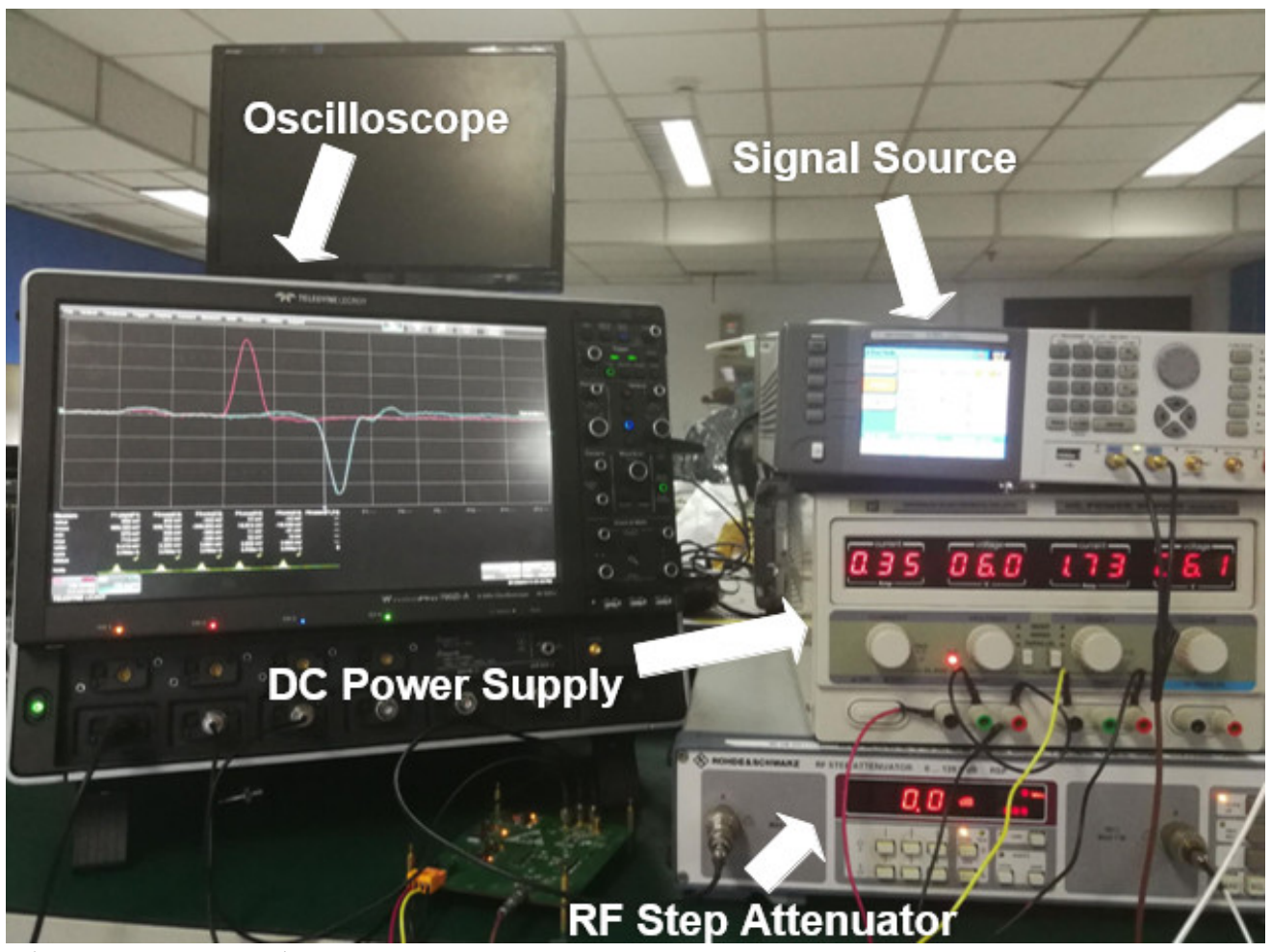

Fig.9 System under test.

A series of tests are carried out to estimate the working condition of electronics.

## B. *Performance Test Results*

### 1) *Functionality Test*

Fig. 10 illustrates that the proposed switch electronics work properly at the switching time of 6 ns. The bipolar signal is successfully divided into two unipolar signals.

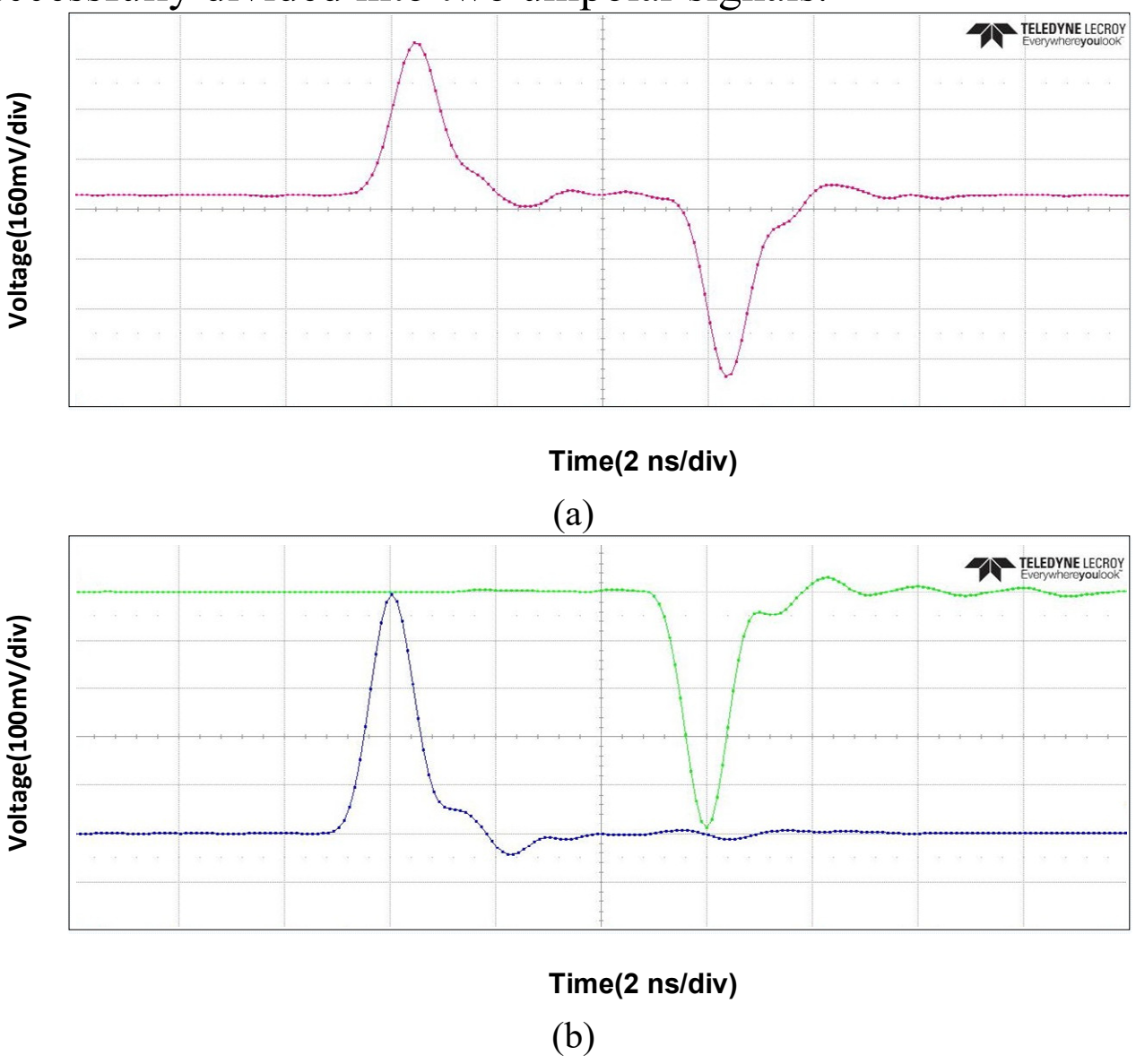

Fig. 10 Waveforms at the switching time of 6 ns. The bipolar signal (a) is successfully divided into two unipolar output signals (b).

In order to explore its functionality under higher speed, tests are conducted under the switching time of 3.6 ns. The results show that the proposed setup can meet the requirements, as shown in Fig. 11.

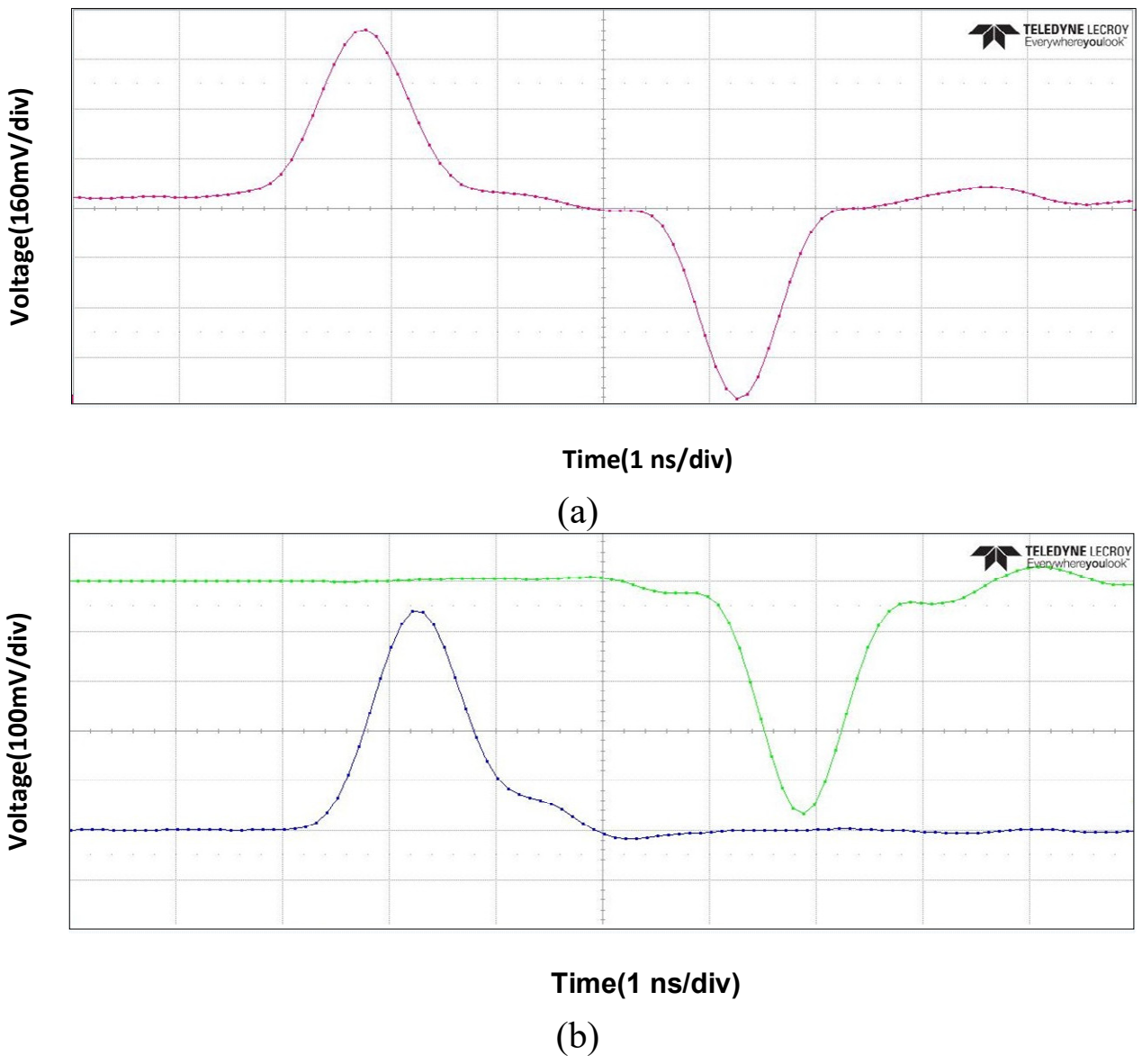

Fig.11 Waveforms at the switching time of 3.6 ns. The bipolar signal (a) is successfully divided into two unipolar output signals (b).

In order to further explore the performance of proposed circuit, several performance parameters are measured.

### 2) *Insertion Loss Test*

Insertion loss describes the loss of signal power, which originates from the electronics. Given the input and output signals, the loss of signal power resulting from the switch electronics is calculated using Eq. (1).

$$InsertionLoss = 20\log\frac{V_{out}}{V_{in}} \quad (1)$$

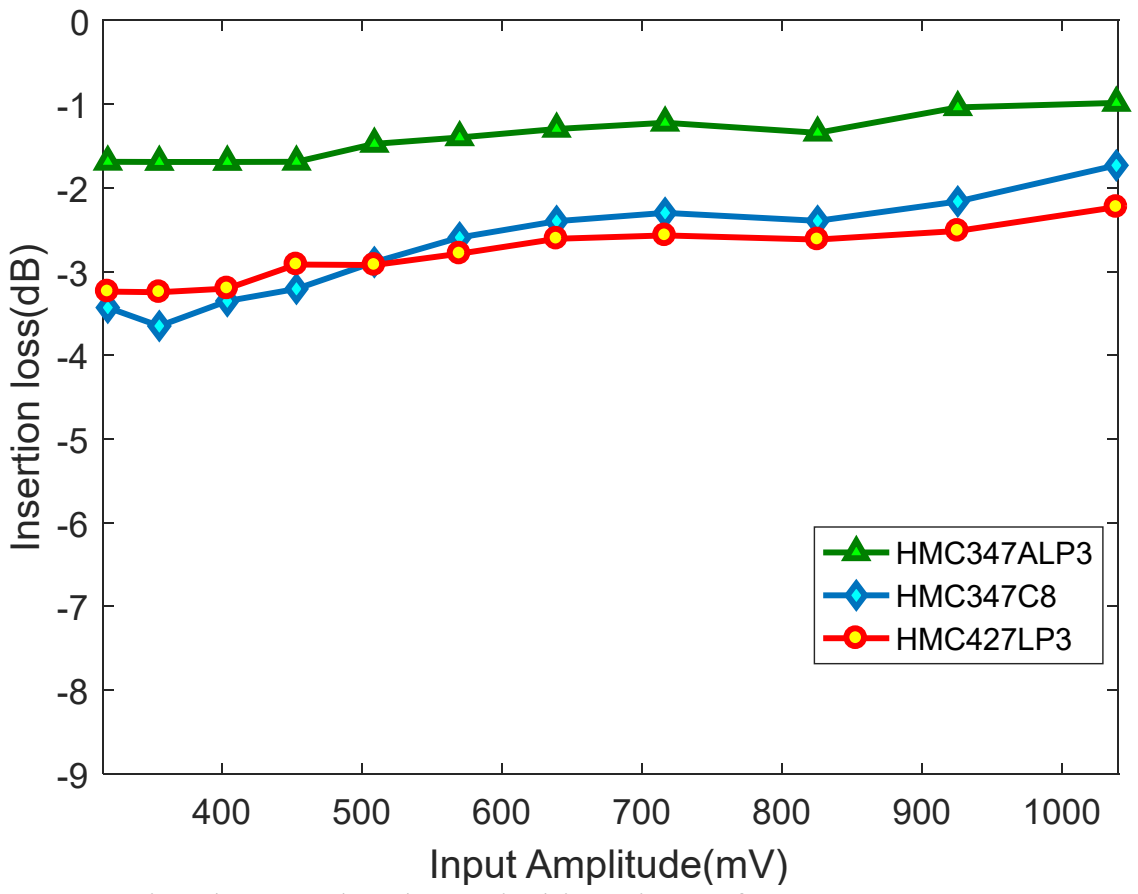

Fig.12 Insertion loss under the switching time of 6 ns.

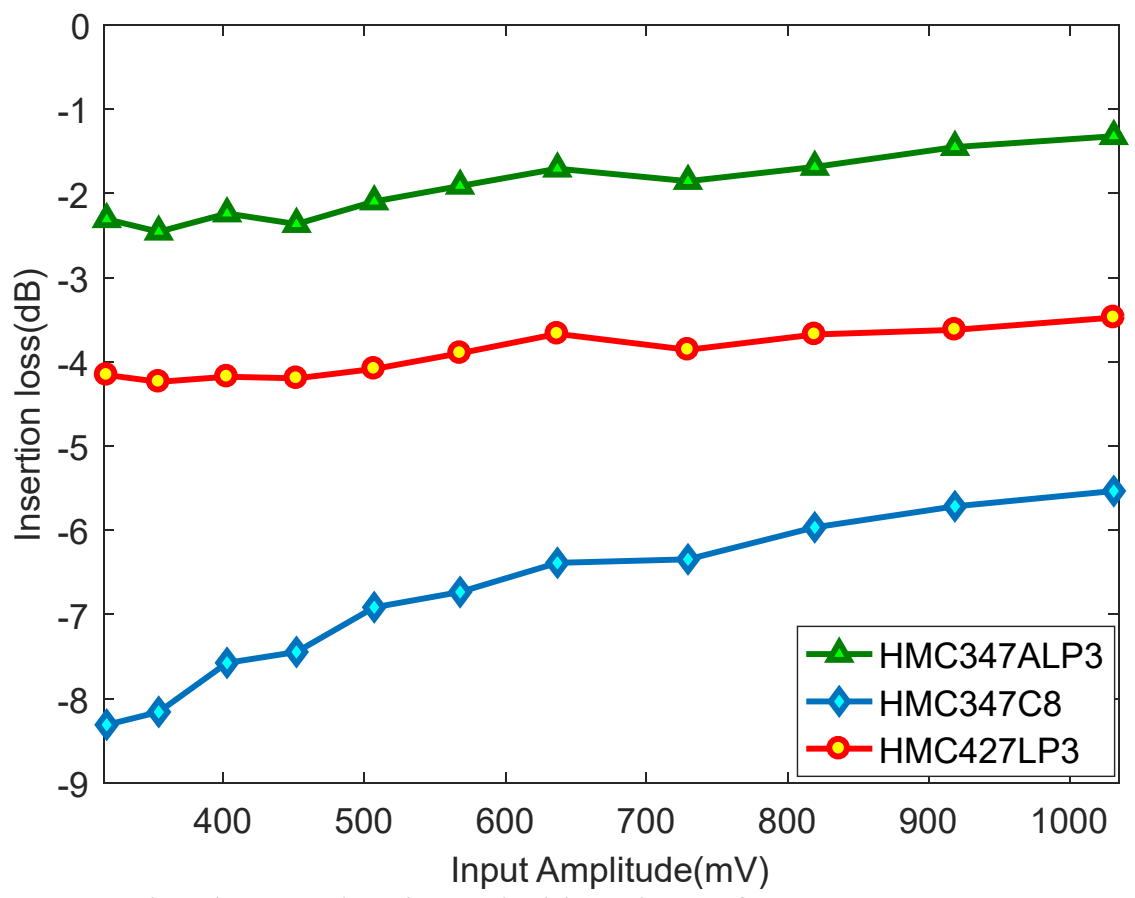

Fig.13 Insertion loss under the switching time of 3.6 ns.

When the RF switch is switching, the insertion loss is expected to be better than -2 dB. Fig. 12 and Fig. 13 suggest that the circuit has exceeded the expectation when switching time is set to be 6 ns. HMC347ALP3 performs well when the bipolar signal has a high amplitude, which is consistent with the simulation results.

In order to explore its limit, the switching time is then set to be 3.6 ns. The results show that a particular version of the switch electronics has the potential to meet the higher speed working requirements.

3) *Isolation Test*

Isolation is a criterion describing how well the input and output ports are disconnected when the RF switch is not switching. When the network analyzer is put into use, the isolation of the switch can be conveniently measured.

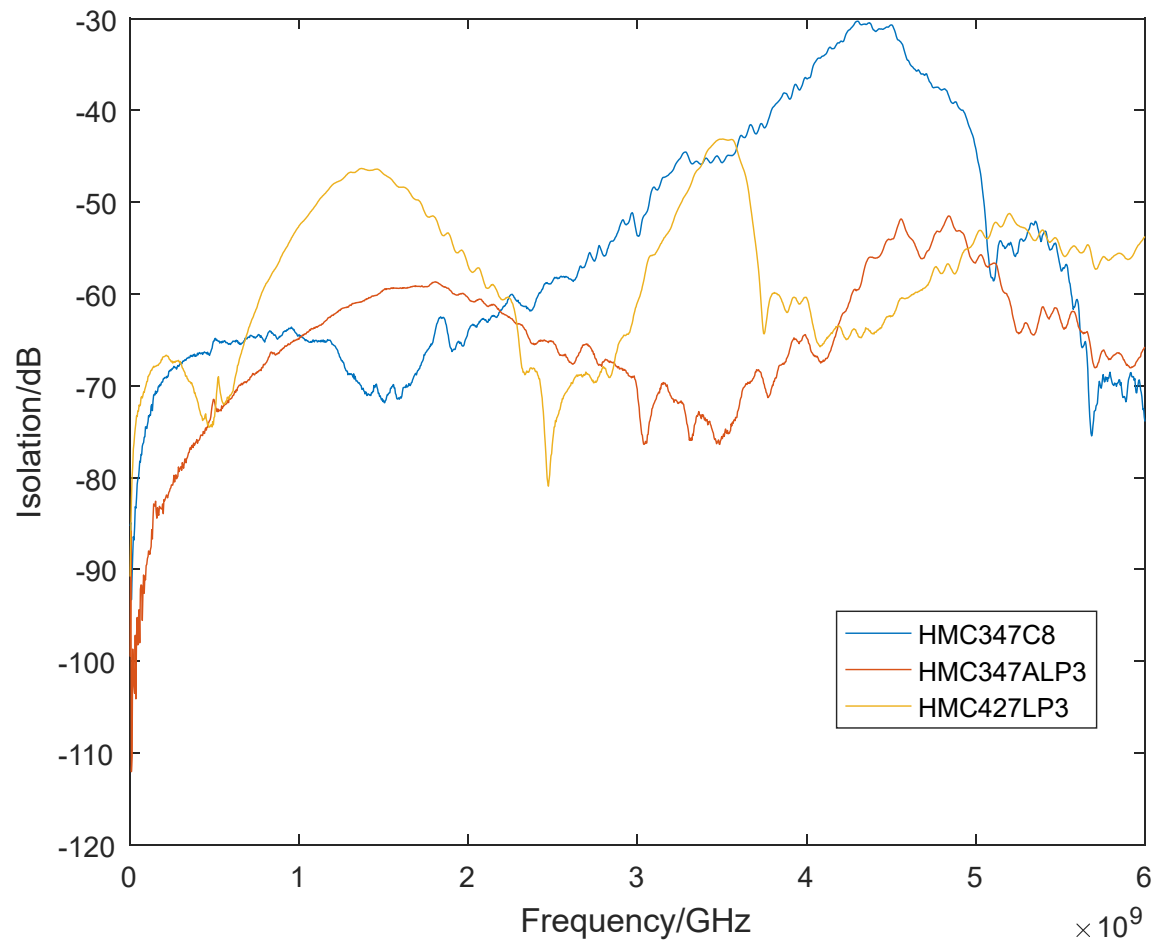

Fig. 14 Isolation test in the frequency domain.

Fig. 14 shows the results of isolation tests in the frequency domain. The results suggest that the isolation of RF switch electronics equipped with HMC347ALP3 and HMC427ALP3 is better than -35 dB when the frequency is under 6 GHz. The test results are consistent with the simulation results.

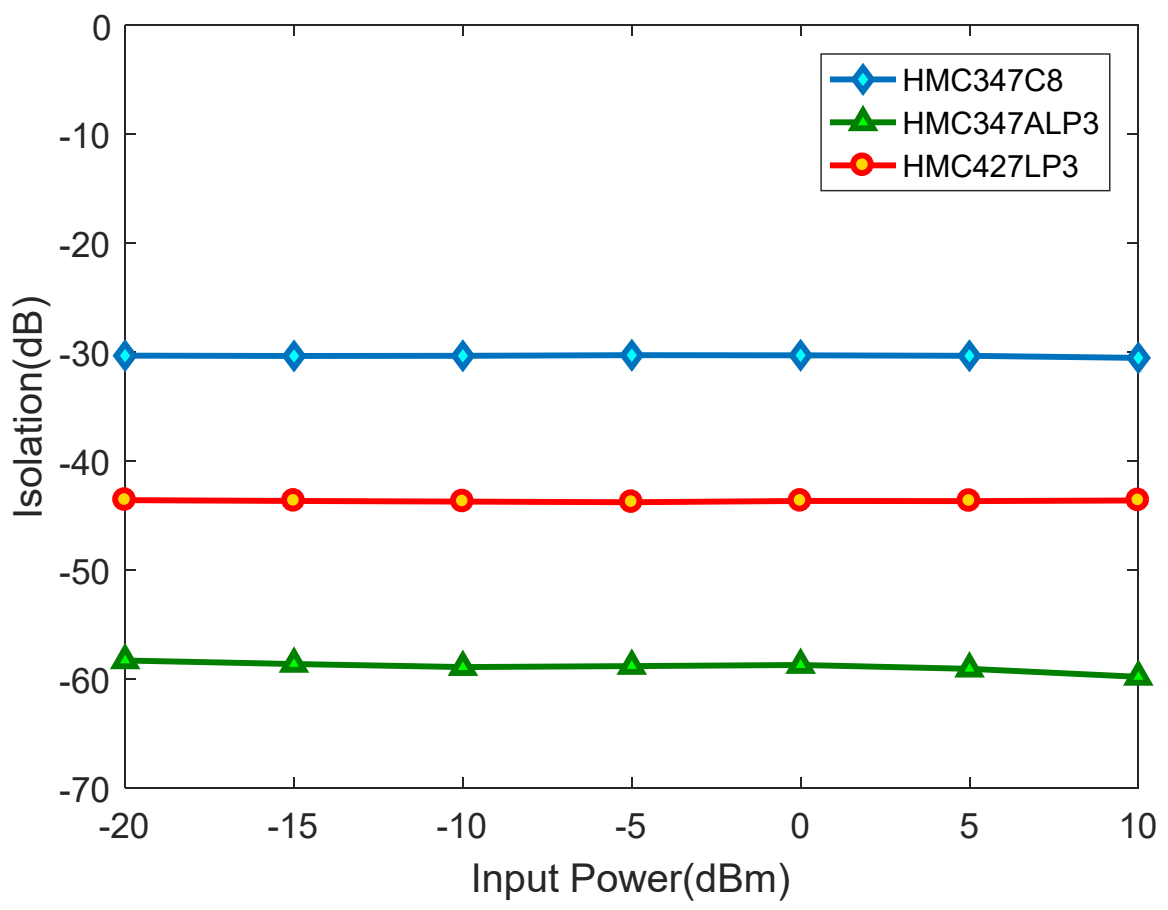

Fig. 15 Isolation measurement results.

Fig. 15 shows that the input and output ports are well disconnected when the RF switch electronics is in non-switching state. This guarantees the reliability of the proposed system.

4) *Delay Line Performance Test*

In order to synchronize the differential signal pair, the RF switch circuit employs delay line to precisely adjust the switching moment. When the TTL controlling signals are not precisely aligned, both the electron bunch and the positron bunch will be distorted. In order to explore this phenomenon, a series of tests have been conducted.

The delay line circuit adopts dial switch to adjust the delay time. The calibration of the delay line circuit has been conducted, and the corresponding results are shown in Fig. 16.

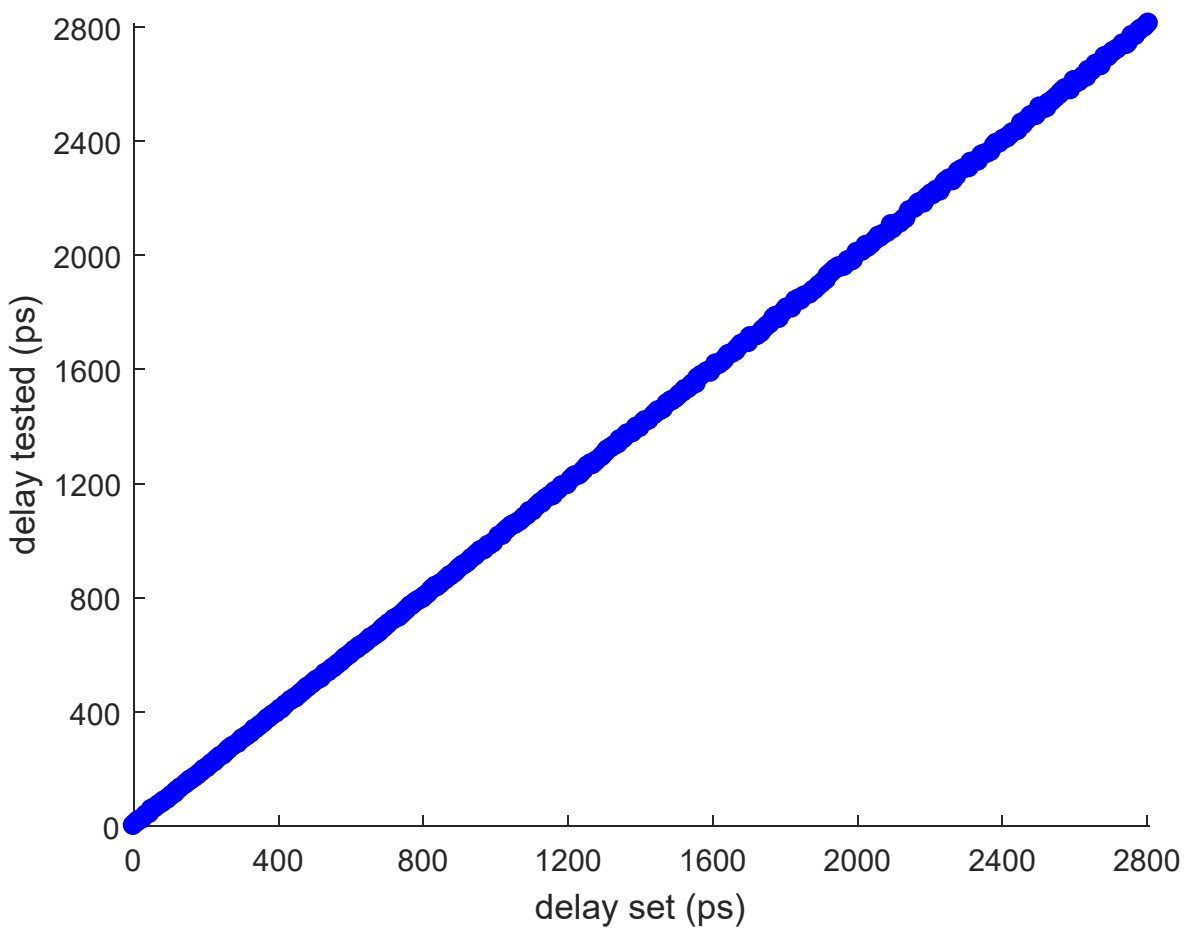

Fig. 16 Calibration curve of the delay line circuit.

As mentioned earlier, the waveform is closely related to the skew of the TTL controlling signals, which is described as the relative delay time. In this work, the ratio of amplitude of two unipolar signals (Vp/Vn) is used to describe the distortion of waveform. Then, the delay time of one of the TTL controlling signals is fixed. By adjusting the delay time of the other TTL controlling signal, the relationship between the *Vp/Vn* ratio and the relative delay time is explored. The delay time is calibrated with a step size of 10 ps for high precision.

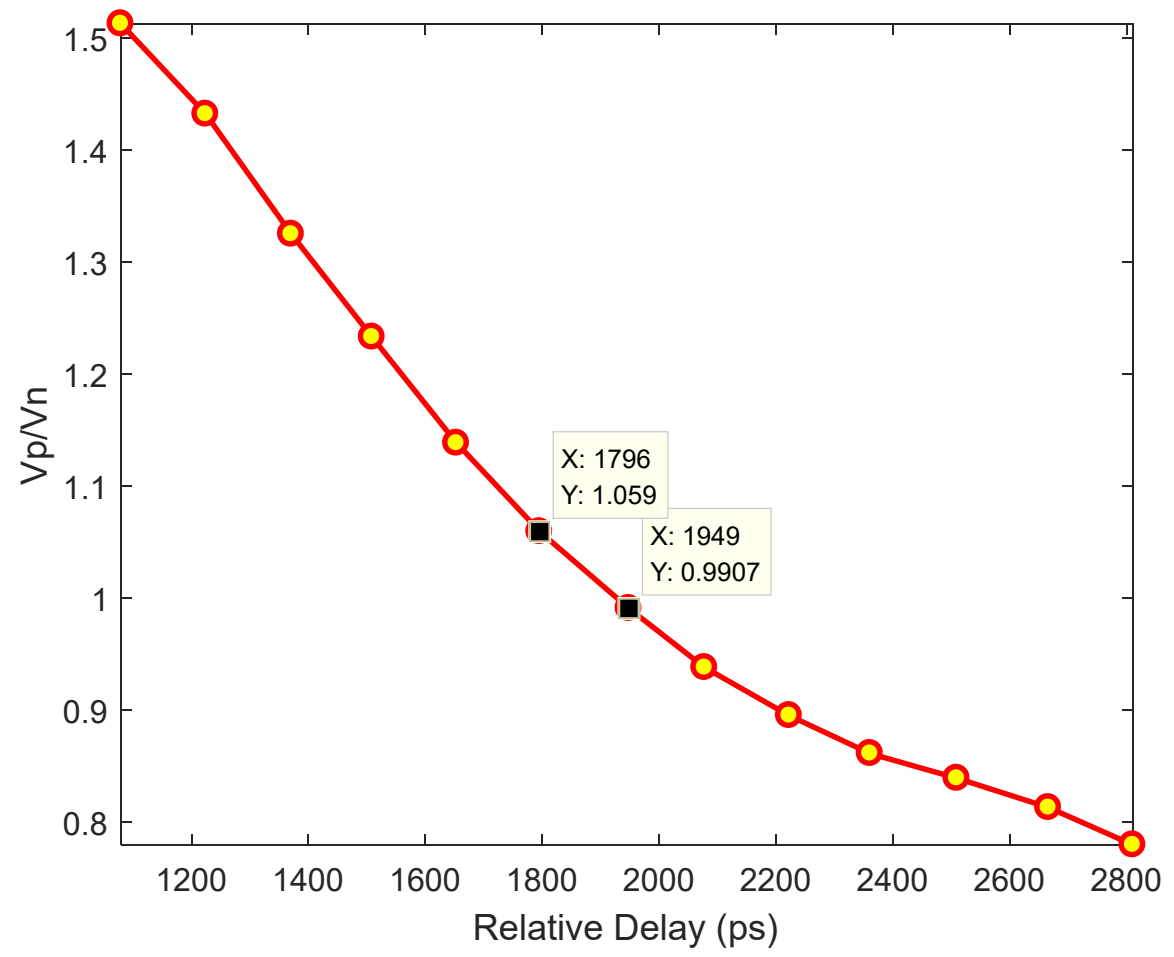

Fig. 17 Relationship between the relative delay time and the signal amplitude ratio.

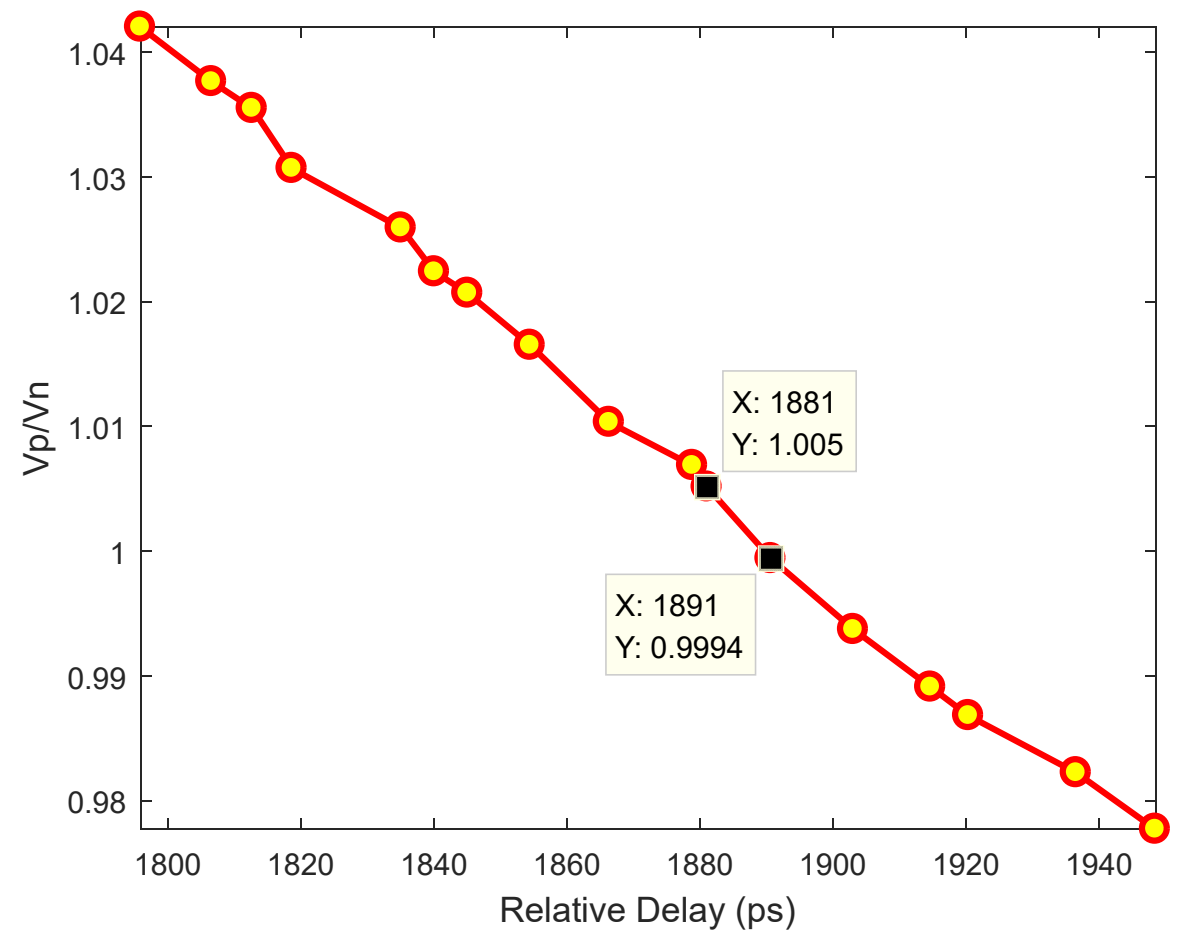

Fig. 18 Optimal relative delay adjustment under the switching time.

As shown in Fig. 17, the waveform is least distorted when the *V*p/*V*n ratio is closest to unity. If the relative delay time is too high or too low, the waveform will be distorted to a certain extent. In order to reach the optimal relative delay time, a series of fine tuning attempts are made. The corresponding results are shown in Fig. 18. The optimal delay adjustment mentions that a relative delay time between 1881 ps and 1891 ps will lead to a low distortion.

In the practical picking up of the beam bunches, however, not only should the two TTL controlling signals be precisely aligned, but the switching moment marked by the aligned TTL signals should be finely adjusted as well. In other words, both the relative delay time and the common delay time should be precisely set to reach the least distortion. To explore the impact of both factor on the *Vp/Vn* ratio, a series of test has been conducted. The results are shown in Fig. 19 and Fig. 20.

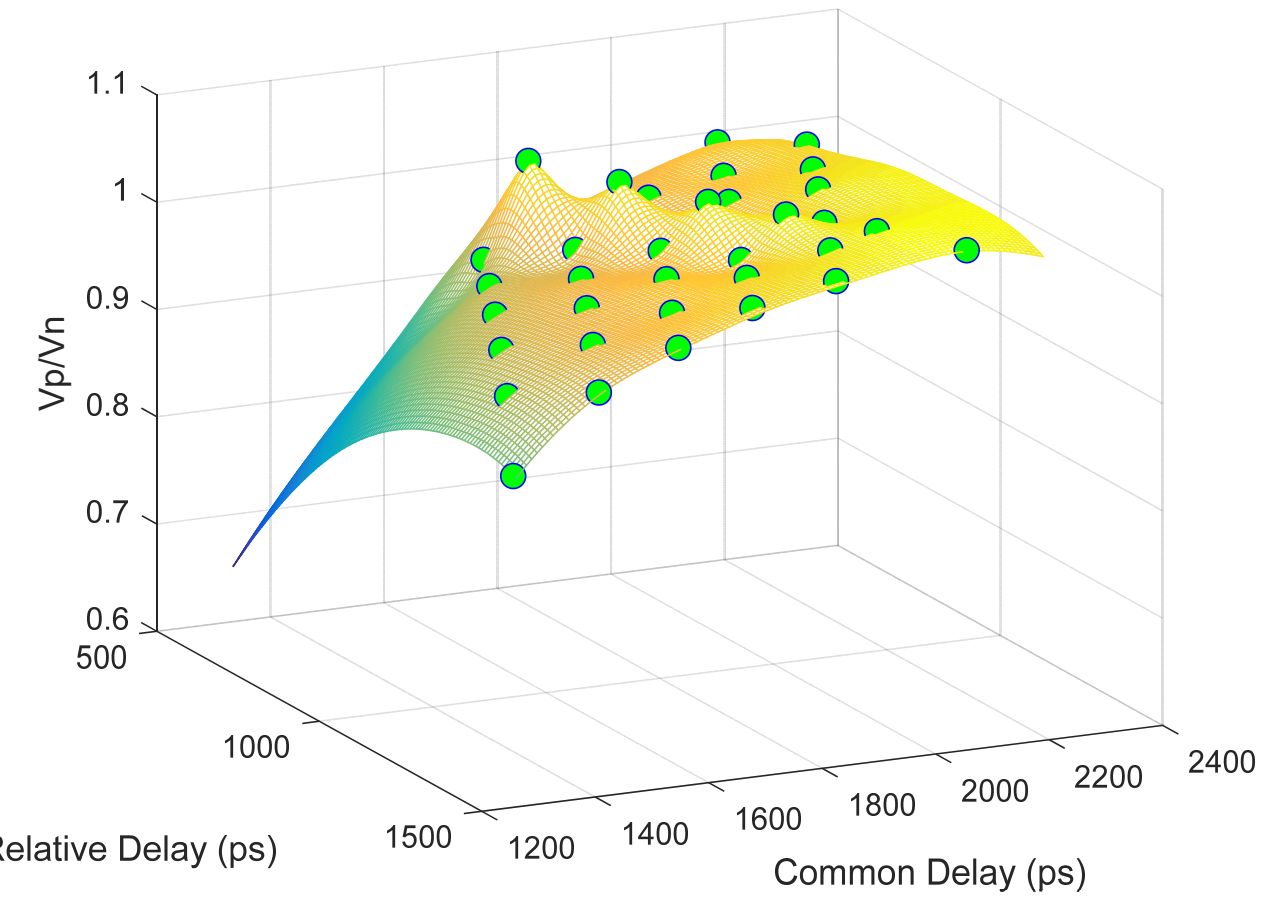

Fig. 20 Relationship between the delay time and the signal amplitude ratio.

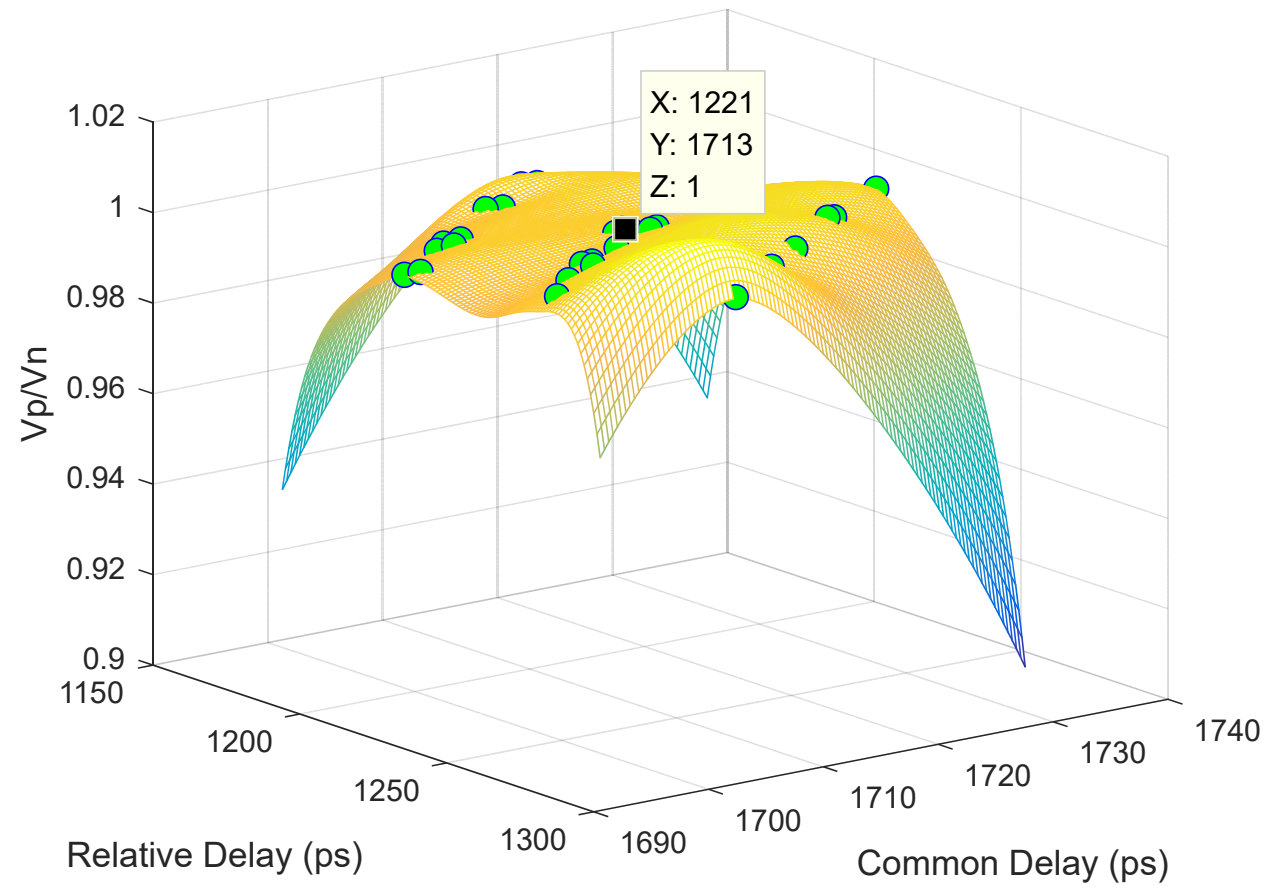

Fig. 21 Optimal delay adjustment under the switching time

The results suggest that the relative delay time should be between 1206 ps and 1223 ps, and the common delay time should be between 1711 ps and 1722 ps to get least distortion. Through the adjustment of 10 ps step size, an error of 1% is achieved.

## IV. Conclusions

High-speed RF switch electronics is designed for picking-up of electron-positron bunches. Initial test results have revealed that the circuit achieves a switching time of less than 6 ns, an insertion loss of better than -2 dB, and an isolation of -55 dB, which provide important foundations for further research in this field. Moreover, the switching moment can be precisely set by adjusting the delay time. The optimal delay adjustment with a step size of 10 ps has been conducted to reach an error of 1%.

## Acknowledgment

The authors would like to appreciate Dr. Huizhou Ma and all other collaborators in IHEP, CAS for their kindly help.

## REFERENCES

[1] J. Power, M. Stettler, The Design and Initial Testing of a Beam Phase and Energy Measurement for LEDA, Proceeding of the 1998 Beam Instrumentation Workshop (BIW98) conference, 1978.

[2] J. Y. Huang, D. H. Jung, D. T. Kim et al, "PLS Beam Position Monitoring System", 3rd European Particle Accelerator Conference, Berlin, Germany, 1992, p. 1115-1117.

[3] J. Power and M. Stettler, "The design and initial testing of a beam phase and energy measurement for LEDA," in *Proc. AIP Conf.*, 1998, vol.451, pp. 459–466.

[4] M. Geitz, "Bunch Length Measurements," *Proceedings DIPAC* 1999 – Chester, UK.

[5] L. Zhao, S. Liu, and X. Gao, "Beam Position and Phase Measurement System for the Proton Accelerator in ADS," *IEEE Transactions on Nuclear Science,* vol. 61, No.1, Feb 2014.

[6] D Olsson, L Malmgren, A Karlsson, "The Bunch-by-Bunch Feedback System in the MAX IV 3 GeV Ring," (2017) In Technical Report LUTEDX/(TEAT-7253)/1-48/(2017) 7253.

[7] C.X. Yin, D.K. Liu, L.Y. Yu, Digital Phase Control System for SSRF LINAC, Proceedings of ICALEPCS07, 2007, p.717-719.

[8] G. Oxoby, R. Claim, "Bunch-by-Bunch Longitudinal Feedback System for PEP-II", presented at the Fourth European Panicle Accelerator Conference (EPAC94). London, England. June 27-My 1.1994.

[9] Jinxin Liu, Lei Zhao, "Bunch-by-Bunch Beam Transverse Feedback Electronics Designed for SSRF", IEEE TRANSACTIONS ON NUCLEAR SCIENCE, VOL. 64, NO. 6, JUNE 2017.

[10] Bernardini C, "AdA: The first electron-positron collider". Phys Persp, 2004, 6: 156–183.

[11] Hübner K, "Designing and building LEP". Phys Rep, 2004, 403-404: 177–188 44.

[12] Aßmann R. "A brief history of the LEP collider" Nucl Phys B Proc Suppl, 2002, 109: 17–31 45.

[13] European Organization for Nuclear Research. "LEP design report: Vol II the LEP main ring", CERN-LEP/84-01, June 1984.

[14] Blondel A P, Geneva U, Zimmermann F, "LEP3: A high luminosity e+ e− collider in the LHC tunnel to study the Higgs boson". Proceedings of IPAC 2012, New Orleans, Louisiana, USA, TUPPR078.

[15] Rees J R. "The positron-electron project-PEP", SLAC –PUB -1911/PEP-243, March 1977

[16] H. Yan, L. Zhao and S. Liu "A beam position measurement system of fully digital signal processing at SSRF," *Nucl.Sci.Technol.*, vol. 23, no. 2, pp. 75–82, 2012.

[17] M. Tobiyama and E. Kikutani, "Development of a high-speed digital signal process system for bunch-by-bunch feedback systems," *Phys. Rev.Special Topics-Accel. Beams*, vol. 3, no. 1, p. 012801, Jan. 2000

[18] Y. Wang, (2014, Feb 13). Networks [Online]. Available: https://indico.cern.ch/event/282344/contributions/1630738/attachments/519381/716573/7-yifang-CEPC-SppC.pdf.

[19] HMC427LP3 datasheet, Analog Device Inc. [Online]. Available: http://www.analog.com/media/en/technical-documentation/data-sheets/hmc427.pdf.

[20] HMC347C8 datasheet, Analog Device Inc. [Online]. Available: http://www.analog.com/media/en/technical-documentation/data-sheets/hmc347c8.pdf.

[21] HMC347ALP3 datasheet, Analog Device Inc. [Online]. Available: http://www.analog.com/media/en/technical-documentation/data-sheets/HMC347ALP3E.pdf.